BuRNN: Buffer Region Neural Network Approach for Polarizable-Embedding Neural Network / Molecular Mechanics Simulations

*Bettina Lier,[#a] Peter Poliak,[#a,b] Philipp Marquetand,[c] Julia Westermayr[*d] and Chris Oostenbrink[*a]*

a.	Institute for Molecular Modeling and Simulation, Department of Material Sciences and Process Engineering, University of Natural Resources and Life Sciences, Vienna. Muthgasse 18, 1190 Vienna Austria

b.	Department of Chemical Physics, Institute of Physical Chemistry and Chemical Physics, Faculty of Chemical and Food Technology, Slovak University of Technology in Bratislava, Radlinského 9, 812 37 Bratislava, Slovakia

c.	Institute of Theoretical Chemistry, University of Vienna, Währingerstraße 17, 1090 Vienna, Austria

d.	Department of Chemistry, University of Warwick, Gibbet Hill Road, CV4 7AL Coventry, UK.

*# Authors contributed equally*

*\* Corresponding authors: julia.westermayr@warwick.ac.uk and chris.oostenbrink@boku.ac.at*

## ABSTRACT

Hybrid quantum mechanics/molecular mechanics (QM/MM) simulations have advanced the field of computational chemistry tremendously. However, they require the partitioning of a system into two different regions that are treated at different levels of theory, which can cause artefacts at the interface. Furthermore, they are still limited by high computational costs of quantum chemical calculations. In this work, we develop BuRNN, an alternative approach to existing QM/MM schemes, which introduces a buffer region that experiences full electronic polarization by the inner QM region to minimize artefacts. The interactions between the QM and the buffer region are described by deep neural networks (NNs), which leads to high computational efficiency of this hybrid NN/MM scheme while retaining quantum chemical accuracy. We demonstrate the BuRNN approach by performing NN/MM simulations of the hexa-aqua iron complex.

## TOC GRAPHICS

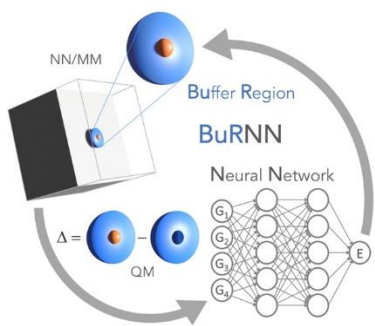

Molecular dynamics (MD) simulations are powerful tools to study the dynamics of systems consisting of hundreds of thousands of atoms. The energy of the system can be described fully classically by a molecular mechanics (MM) force field, by a quantum mechanical (QM) method, or by a hybrid quantum mechanics/molecular mechanics (QM/MM) technique. The latter approach is very powerful, as it enables an accurate description of a small important part of a system at the appropriate level of quantum chemistry, while the remainder is treated by MM to simulate large system sizes at relevant time scales.[1]

In QM/MM approaches, the electrostatic coupling between the partitioned regions can be treated with different levels of mutual interaction, i.e., embedding schemes.[2-5] Mechanical embedding is the simplest and least accurate approach. Interactions are described via classical point charges only. In contrast, electrostatic embedding is physically better motivated, as the QM system experiences the MM charge distribution being embedded in the QM Hamiltonian. However, QM particles see MM particles as fixed point charges, which neglects polarization in the MM region. To account for polarization effects in the MM region as well, polarizable force fields can be used.[6,7]

Independently of the scheme, all QM/MM methods are limited by high computational costs of the quantum calculation and issues at the interface, such as overpolarization.[8] Particularly prone to such artifacts are boundaries that cross covalent bonds, although a careful choice of the bond splitting scheme can alleviate them.[4] Furthermore discrepancies between the forces derived for the QM and MM region can lead to artifical crowding or depletion at the interface, when particles are allowed to change character during a simulation. Several approaches have been proposed to address boundary artifacts either by introducing an intermediate region[9,10] or by restricting the boundary transition.[11]

Alternatively, the whole system can be treated using machine-learned interatomic potentials based on ab initio data.[12] Machine learning (ML) is especially effective for MD simulations as it can learn the relation between a descriptor, i.e., the structure of a system, and a targeted output, i.e., energies and forces with the accuracy of the reference method, but much lower computational costs. Such ML potentials are available for specific materials at different levels of theory.[13-17] However, universal ML potentials for more complex systems, such as biomolecules, still pose a challenge and are limited by the computational expenses of the reference calculations.[2, 12]

Very recently, ML potentials have been combined with QM/MM concepts and were shown to be powerful to, e.g., calculate free energies or transition paths.[18-24] However, these approaches are complicated as ML models need to capture the effects of the environment (MM region) even though only the QM region has to be learned. The introduction of a cut-off, up to which the MM region is included, has emerged as a solution.[21, 22, 25]. One example is FieldSchNet,[20] which circumvents this problem by sampling the environment while keeping the QM region fixed. This model has been shown powerful to predict spectra and chemical reactions with neural networks (NNs) using electrostatic embedding, but requires extended sampling. Due to the nature of electrostatic, artefacts at the interfaces are not reduced in aforementioned approaches.

To circumvent boundary problems and with the aim to avoid extensive force field parameterizations we propose an alternative approach. We introduce an additional buffer region, that experiences full electronic polarization by the inner QM region. The buffer region is described at both the QM and the MM level. Effectively, the interactions with the QM region are calculated entirely at the QM level while the interactions with the MM region are described at the MM level. Within the buffer region, the interactions are a combination of MM interactions and the effect of the QM region on the electronic degrees of freedom of the buffer region. While this approach minimizes the artefacts that arise from mixing two levels of theory, it comes at considerable computational costs as two QM calculations are required. By using ML to describe QM-derived energy surfaces, an elegant solution emerges. In this work, we train deep NNs to directly predict the difference of the two required QM calculations. Thus, this scheme automatically includes the mutual influence of the QM and the buffer region, without the need for additional external forces in the ML setup.[20] The calculation of all relevant energies in a simulation can efficiently be done with a single evaluation of the NN. Due to the fact that interaction energies are often easier to learn with NNs than potential energies, outstanding accuracy can be achieved with mean absolute errors in the range of a few kJ/mol. This range is well below the often-desired chemical accuracy of machine learning models, and is independent of the size of the inner region. We refer to this NN/MM approach as <u>Bu</u>ffer <u>R</u>egion with <u>N</u>eural <u>N</u>etwork, or short BuRNN. Although schemes like ONIOM with different regions exist, this approach is, to the best of our knowledge, novel.

The BuRNN approach partitions a system into an inner region $\mathbb{I}$, a buffer region $\mathbb{B}$, and an outer region $\mathbb{O}$ (Figure 1). The total potential energy, $V_{tot}$, contains the QM energy of the inner and buffer region $V_{\mathbb{I}+\mathbb{B}}^{QM}$ and the MM energy coupling of the outer region to all other regions, $V_{\mathbb{O}(\mathbb{I}+\mathbb{B}+\mathbb{O})}^{MM}$. The buffer region is calculated at both levels of theory. The difference of the two buffer terms $V_{\mathbb{B}}^{MM} - V_{\mathbb{B}}^{QM}$ is included in $V_{tot}$ and helps to smoothen the transition between the QM energy of the

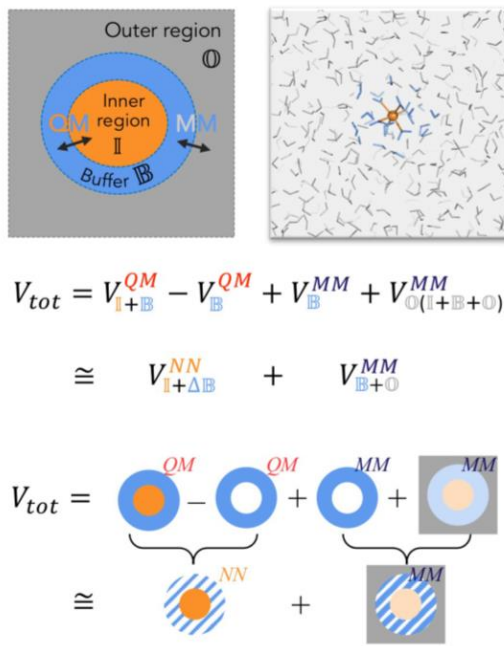

$$V_{tot} = V_{\mathbb{I}+\mathbb{B}}^{QM} - V_{\mathbb{B}}^{QM} + V_{\mathbb{B}}^{MM} + V_{\mathbb{O}(\mathbb{I}+\mathbb{B}+\mathbb{O})}^{MM}$$
$$\cong V_{\mathbb{I}+\Delta\mathbb{B}}^{NN} + V_{\mathbb{B}+\mathbb{O}}^{MM}$$

Figure 1: Scheme of the BuRNN approach, which distinguishes three regions: 1) the inner region $\mathbb{I}$ (orange), which is described entirely by quantum mechanics (QM); 2) the buffer region $\mathbb{B}$ (blue), which is described both by QM and molecular mechanics (MM) and 3) the outer region $\mathbb{O}$ (grey) which is described entirely by a classical MM force field.

inner region and MM energy of the outer region. In addition, artefacts in the electronic degrees of freedom at the outer edge of the buffer region will largely cancel in the difference $V_{\mathbb{I}+\mathbb{B}}^{QM} - V_{\mathbb{B}}^{QM}$. Further details are discussed below. Adding all terms for the total potential energy together leads to:

$$V_{tot} = V_{\mathbb{I}+\mathbb{B}}^{QM} - V_{\mathbb{B}}^{QM} + V_{\mathbb{B}}^{MM} + V_{\mathbb{O}(\mathbb{I}+\mathbb{B}+\mathbb{O})}^{MM}. \qquad 1$$

The subscripts for the potential energy denote the calculated region, the superscripts the method. Even though interactions in QM are not pairwise additive, it is instructive to consider them as hypothetical pairwise interactions within or between regions. They are indicated with a comma-separated subscript. $V_{\mathbb{I}+\mathbb{B}}^{QM}$ can then be separated into three terms, i.e., the energy that results from interactions within the inner region, $V_{\mathbb{I},\mathbb{I}}^{QM}$, between the inner and buffer region, $V_{\mathbb{I},\mathbb{B}}^{QM}$, and within the buffer region, $V_{\mathbb{B},\mathbb{B}}^{QM}$:

$$V_{\mathbb{I}+\mathbb{B}}^{QM} = V_{\mathbb{I},\mathbb{I}}^{QM} + V_{\mathbb{I},\mathbb{B}}^{QM} + V_{\mathbb{B},\mathbb{B}}^{QM}. \qquad 2$$

The potential energy of the buffer region, $V_{\mathbb{B}}^{QM}$, is separately calculated at the QM level as well and is denoted as $V_{\mathbb{B},\mathbb{B}}^{QM(isolated)}$. This should emphasize that the inner region is not part of this particular calculation and term. $V_{\mathbb{B},\mathbb{B}}^{QM}$ as hypothetical pairwise interaction within the buffer region as used in equation 2, though, includes the influence of the inner region on the buffer. The difference of the two terms can be seen as the polarization of the buffer, which is denoted as $V_{\mathbb{B},\mathbb{B}}^{QMpol}$:

$$V_{\mathbb{B},\mathbb{B}}^{QMpol} = V_{\mathbb{B},\mathbb{B}}^{QM} - V_{\mathbb{B},\mathbb{B}}^{QM(isolated)} \qquad 3$$

The energy of the buffer is also described at the MM level, $V_{\mathbb{B}}^{MM} = V_{\mathbb{B},\mathbb{B}}^{MM}$. If this term agrees with QM exactly, it will cancel with $V_{\mathbb{B},\mathbb{B}}^{QM(isolated)}$. The outer region with all involving interactions $V_{\mathbb{O},\mathbb{O}}^{MM}$, $V_{\mathbb{I},\mathbb{O}}^{MM}$ and $V_{\mathbb{B},\mathbb{O}}^{MM}$, are treated at the MM level, but with partial charges of the inner and buffer regions derived from the QM calculation of inner and buffer region together. Hence it also includes a representation of the polarization of the inner and buffer regions:

$$V_{\mathbb{O}(\mathbb{I}+\mathbb{B}+\mathbb{O})}^{MM} = V_{\mathbb{I},\mathbb{O}}^{MM} + V_{\mathbb{B},\mathbb{O}}^{MM} + V_{\mathbb{O},\mathbb{O}}^{MM} \qquad 4$$

The total energy in terms of hypothetical interaction energies can then be written as:

$$V_{tot} = V_{\mathbb{I},\mathbb{I}}^{QM} + V_{\mathbb{I},\mathbb{B}}^{QM} + V_{\mathbb{I},\mathbb{O}}^{MM} + V_{\mathbb{B},\mathbb{B}}^{QMpol} + V_{\mathbb{B},\mathbb{B}}^{MM} + V_{\mathbb{B},\mathbb{O}}^{MM} + V_{\mathbb{O},\mathbb{O}}^{MM} \qquad 5$$

Thus, BuRNN ensures that the interactions within and between neighboring regions are computed at the appropriate levels. One of the benefits is that any artefacts in the electronic degrees of freedom will cancel in the difference $V_{\mathbb{I}+\mathbb{B}}^{QM} - V_{\mathbb{B}}^{QM}$, as mentioned above. This is true with the assumption that these artefacts are due to the interface to the outer region and that the relevant polarization of the buffer region predominantly takes place at the interface between the inner and the buffer region. More specifically, artefacts in the QM calculation may arise in the electron density at the outer boundary of the buffer region, which differs from a solvated boundary. However, these artifacts will cancel in the difference as it will be very similar in both QM terms. Any remaining artefacts potentially arising at the interface between the buffer and the outer region are, furthermore, relatively far away from the inner region of interest.

The interactions between $\mathbb{I},\mathbb{O}$ as well as $\mathbb{B},\mathbb{O}$ are computed using a mechanical embedding scheme with charges assigned from the QM calculation, which is appropriate because of the large distances between inner and outer regions. Direct electronic influences on the inner region due to the outer region will be relatively small and the interaction is largely electrostatic. Remaining artefacts may arise from the fact that particles moving from the outer region into the buffer region switch from a mechanically embedded MM interaction to a full QM interaction with the inner region. Also in this case, the interaction is at a relatively long distance from the inner region, where the $\mathbb{I}, \mathbb{B}$ interaction will be largely of electrostatic nature, such that these artefacts can be expected to be small.

Despite the accuracy and benefit of this scheme, the burden lies in the high computational costs that remain because two computationally expensive QM calculations are required. To overcome this limitation, we describe the first two terms of equation 1 directly using a deep NN:

$$V_{\mathbb{I}+\Delta\mathbb{B}}^{NN} \cong V_{\mathbb{I}+\mathbb{B}}^{QM} - V_{\mathbb{B}}^{QM}, \qquad 6$$

which is equal to

$$V_{\mathbb{I}+\Delta\mathbb{B}}^{NN} = V_{\mathbb{I},\mathbb{I}}^{QM} + V_{\mathbb{I},\mathbb{B}}^{QM} + V_{\mathbb{B},\mathbb{B}}^{QMpol}. \qquad 7$$

It now becomes clear that the introduction of a buffer region in BuRNN is akin to a polarizable embedding with the polarization described at the full QM level. Thus, the NN represents the full interactions within the inner region, the interactions between the inner and buffer region, and the polarization of the buffer region due to the inner region in a single term, $V_{\mathbb{I}+\Delta\mathbb{B}}^{NN}$. The delta-sign is used to emphasize that BuRNN essentially includes a delta-learning,[26] bringing interactions of the buffer region from the MM to the QM level. The total BuRNN energy can finally be rewritten to:

$$V_{tot} = V_{\mathbb{I}+\Delta\mathbb{B}}^{NN} + V_{\mathbb{I},\mathbb{O}}^{MM} + V_{\mathbb{B},\mathbb{B}}^{MM} + V_{\mathbb{B},\mathbb{O}}^{MM} + V_{\mathbb{O},\mathbb{O}}^{MM}. \qquad 8$$

All MM terms can be computed classically from a single call to the force field. The workflow of an NN/MM BuRNN simulation is illustrated in Figure 2. The training data set is based on QM calculations and can be generated via sampling from MD simulation snapshots of the targeted system and extended using adaptive sampling.[27] The training set generation and sampling of initial data is explained in detail in the supporting information sections S1.1 and S1.2. We employ NNs to predict interaction energies, interaction forces, and charges to carry out NN/MM simulations. As NN models we use SchNet,[28, 29] a deep convolutional continuous-filter NN, that was adapted to allow for charge fitting. A full description of the NN models including learning curves, model accuracy, and hyperparameter optimization can be found in section S1.3. The mean absolute error assessed from 5 independently trained NN models is 1.7 ± 0.3 kJ/mol for energies, 8.4 ± 0.4 kJ/mol/nm for forces, and 0.027± 0.001 a.u. for partial charges. Models trained on a larger inner region are comparable in accuracy, as the NN models are always trained on the whole system, i.e., the inner and the buffer region, which are above 50 atoms. For all the outputs, these are very low errors, well below chemical accuracy defined as 1 kcal/mol in recent (machine learning) studies.[26, 30] We have implemented the BuRNN approach in the GROMOS simulation software[31] (see section S1.3). Importantly, we generate the training set and conduct MD simulations by applying two initially trained NNs, A and B. In adaptive sampling, their prediction differences can be used to assess the reliability of NN models during the MD simulation. Whenever the NN predictions for interaction energies are similar, i.e., when $V_{\mathbb{I}+\Delta\mathbb{B}}^{NN(A)} \approx V_{\mathbb{I}+\Delta\mathbb{B}}^{NN(B)}$, predictions are deemed accurate and the simulation is continued. If predictions start to diverge from each other and exceed a predefined threshold, additional reference QM calculations ($V_{\mathbb{I}+\mathbb{B}}^{QM}$ and $V_{\mathbb{B}}^{QM}$) are performed for relevant configurations and added to the training set. In this work, we carried out 4 rounds of adaptive sampling, i.e., until dynamics could be run up to 1 nanosecond without further interruptions. By replacing QM calculations with NNs during MD simulations, BuRNN reduces the computational costs considerably and enables long time scales while retaining high accuracy.

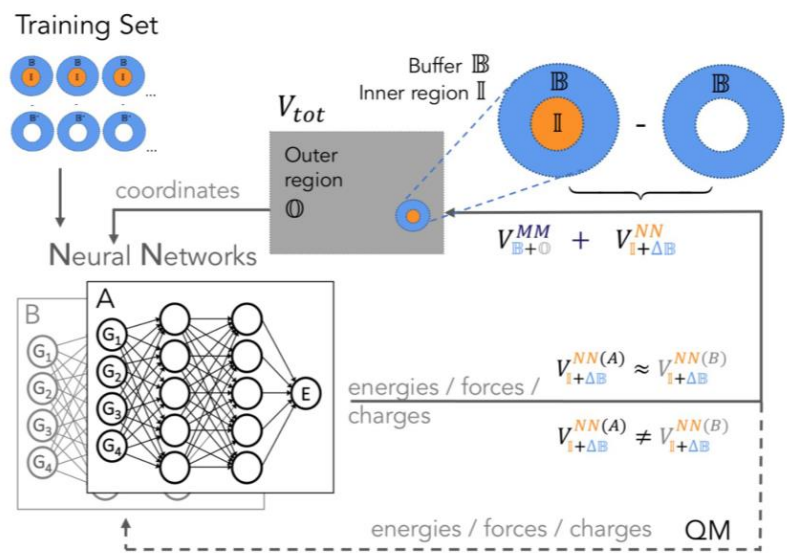

Figure 2: The process of a BuRNN simulation including adaptive sampling. At every $x^{th}$ time step during MM, two neural networks (A and B) are compared. When predictions diverge, the training set is expanded by additional QM calculations.

Here, we demonstrate the use of BuRNN for the hexa-aqua iron [Fe(H$_2$O)$_6$]$^{3+}$ complex in water as a model system. This system has the advantage of being relatively simple to test our approach, but the classical description of transition metal interactions is notoriously difficult, which makes it a good use case of BuRNN for simulating this system. Especially challenging is the coordinative bond between Fe and O as it is somewhat in-between a covalent

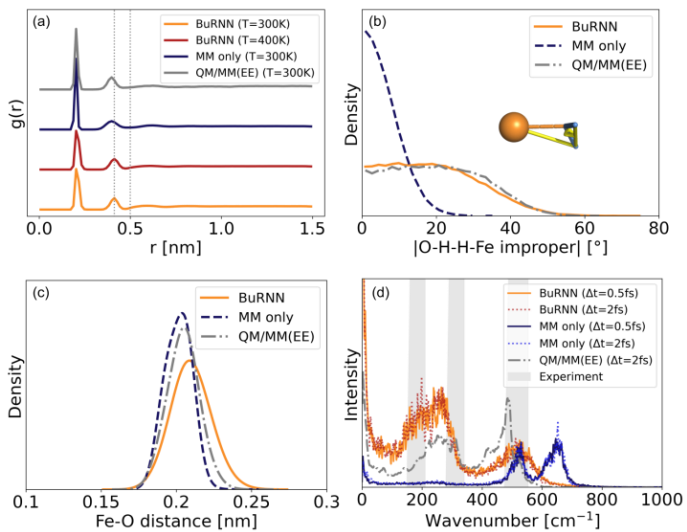

Figure 3: Coordination of Fe$^{3+}$ by water molecules with BuRNN simulations and when using MM only. a) radial distribution function for BuRNN at a temperature of 300K and 400K, with MM only and a QM/MM simulation using electrostatic embedding (EE); dashed lines indicate the second BuRNN peak and the cutoff used to define the buffer region. b) Probability distribution of the O-H-H-Fe improper dihedral, c) distribution of the Fe-O distance, and d) power spectrum of the Fe-O coordinative bond for different simulations. Experimental data are taken from Refs. 50-52.

and an ionic bond and is often addressed with specialized force fields.[32, 33]

In our test case, the $Fe^{3+}$ ion comprises the inner region, water molecules up to 0.5 nm are treated as the buffer region, which roughly accounts for the first two solvation shells, where the first solvation shell is expected to be formed by the hexacoordinated waters. During extensive MD simulations (10 ns), the water molecules are freely diffusing between the buffer and outer regions, smoothly switching interactions between the NN (QM) level and the MM level of theory. BuRNN simulations are validated using experimental data and are compared to QM/MM simulations of the $Fe(H_2O)_6^{3+}$ complex (QM-region) in classical SPC water and two distinct fully classical descriptions. In addition, we executed BuRNN simulations using a larger inner region that additionally comprises the first solvation shell.

To validate the method, we first look at the Fe-O radial distribution functions (RDF) g(r) in Figure 3a. All simulations show a distinct peak at around 0.20 nm with, corresponding to the first coordinative solvation shell of mostly 6 water molecules, which is slightly narrower and more pronounced in the classical description (blue curve). In contrast, the second solvation shell is more pronounced in BuRNN (orange curve) and corresponds to an average of 12.7 water molecules. It shows a maximum at around 0.41 nm, while the MM simulation shows a broader peak with a maximum at 0.40 nm. Simulations with a larger inner region yield almost identical results (Figure S4). In the QM/MM simulation with electrostatic embedding, the second shell also has a maximum at 0.40 nm (grey curve) while a QM/MM simulation with mechanical embedding leads to a maximum at 0.41 nm (Figure S4), as in BuRNN. Experimentally, it was found at 0.415 nm and comprises 12 water molecules, hence agreeing well with our simulations.[34] We have also performed simulations using the 12-6-4 Lennard-Jones potential[35] and the SPC/E water model[36] and find that the g(r) shows an additional peak at 0.31 nm, representing one additional molecule pushing into the first solvation shell (Figure S4). The transition at 0.5 nm in the RDF obtained with BuRNN is smooth and does not show any artifacts. This is remarkable as the buffer region ends and the water molecules beyond this distance interact completely according to a pure MM description.

To investigate the robustness of BuRNN, we performed MD at different temperatures (Figure S5) and show the RDF obtained at 400 K in Figure 3a. As can be seen, there is a slight smoothening between 0.5 and 1.0 nm due to the increased thermal motion, but the BuRNN simulation remains stable. The two NN models deviate on average by 0.39 ± 0.02 kJ/mol.

In addition, we sought to investigate the propensity of BuRNN to describe water exchange. Hence, we used umbrella sampling[37] to pull a water molecule away from the complex and observed the spontaneous exchange of this water molecule by another one (see supporting movie S1). The energy predictions and MD simulations are stable during this process. In the regular simulations, the hexa-coordination is stably maintained. Water molecules in the second solvation shell (within the buffer region) readily exchange with water molecules from the outer region. All water molecules (786 molecules) visit the buffer region at least once during the simulation, with an average lifetime of 14.4 ps. This agrees with estimates from NMR experiments that determine a lifetime which is below their resolution limit of 100 ps.[38] We further computed the self-diffusion rate for BuRNN and MM only simulations and find $0.98 \cdot 10^{-5}$ cm$^2$/ps and $0.92 \cdot 10^{-5}$ cm$^2$/ps, respectively. Both approaches overestimate the diffusion constant compared to experimental estimates of $0.55 - 0.68 \cdot 10^{-5}$ cm$^2$/ps,[38-41] in line with the observation that bulk SPC (simple point charge) water has a slightly too large diffusion constant.[42]

Figure 3b shows the distribution of the O-H-H-Fe improper dihedral angle defining the co-planarity of the iron and a water molecule. A value of 0°, as predominant in pure MM simulations, implies that the water molecule and the $Fe^{3+}$ ion are in the same plane. Larger values as observed for BuRNN with a mean angle of 19.3° and for QM/MM simulations (mean angle of 20.3°) indicate a more tetrahedral arrangement in which the iron interacts with the lone pairs on the oxygen. For comparison, a BP86-D3/def2-TZVP/COSMO estimate for this angle in $[Fe(H_2O)_{18}]^{3+}$ lies at 16°.[43]

MD simulations are further compared by the geometries visited during the simulations. Figures 3c shows radial distances between the Fe-O that agree well with the range of experimental estimates for the Fe-O bond lengths of 0.199 to 0.210 nm.[34, 44-50] O-Fe-O angles are almost identical between BuRNN, QM/MM and "MM only" and reflect angles expected for an octahedral arrangement (peaking at around 90° and 175°). Figure 3d shows that there are clear differences for the frequencies by which the Fe-O bonds vibrate, implying that the Fe-O interaction is indeed not well captured by a purely classical description. In the QM/MM simulations and when using BuRNN the vibrations take place at lower frequencies and are in better agreement with experimental bands observed at ca. 180, 310 and 500 cm$^{-1}$.[50-52] The frequencies obtained with quantum chemistry are better aligned with experiment and BuRNN than with pure MM (Figure S4b) while those obtained from 12-6-4 Lennard Jones potential simulations are even higher than the ones observed with the simple MM only approach (Figure S4).

In this letter, we have introduced BuRNN, a buffered region neural network NN/MM scheme as an alternative to QM/MM simulations that experiences full electronic polarization by the inner QM region. BuRNN minimizes artefacts at the interface between regions by ensuring that interactions that go over boundaries are treated at a consistent level of theory. Inconsistencies at the edges of the buffer i) can be expected to cancel in the difference between two QM terms and ii) are far removed from the inner region. These advantages come at the cost of an additional QM calculation, which is elegantly solved by training NNs directly on the energy difference. A single evaluation of the NN is required to evaluate the energies and a second NN is used to derive charges for full mutual polarization. BuRNN allows fast hybrid NN/MM simulations and has the advantage of being applicable to any system and useable with any molecular ML model.

We have demonstrated the use of BuRNN by realistic simulations of hexa-aqua iron in water. This shows that it is applicable for metal-ligand interactions without the need of additional force field parameters. The good agreement and high stability of BuRNN for long time-scale MD simulations including external perturbation, such as changing temperature or forces that lead to water exchange, make our method very promising for future application of the simulation of more complex systems.

SUPPORTING INFORMATION

Computational details for the quantum calculations, training set generation, training and validation of the neural networks and simulation settings. Further analyses of simulations with alternative settings (QM/MM with mechanical embedding, classical with a 12-6-4 Lennard-Jones potential, BuRNN with alternative charge distributions, BuRNN at higher temperatures, BuRNN with a larger inner region).

AUTHOR INFORMATION

# *Authors contributed equally*

*Corresponding authors: julia.westermayr@warwick.ac.uk and chris.oostenbrink@boku.ac.at*

Author Contributions

Bettina Lier: conceptualization, investigation, data curation, funding acquisition, software, writing - review and editing, visualization; Peter Poliak: investigation, data curation, methodology, software, writing – original draft, review and editing; Philipp Marquetand: methodology, supervision, writing – review and editing; Julia Westermayr: investigation, data curation, methodology, writing – original draft, review and editing, visualization; Chris Oostenbrink: conceptualization, investigation, data curation, methodology, writing – original draft, review and editing, supervision, resources.

Conflicts of interest

The authors declare no competing financial interests.

ACKNOWLEDGMENT

We thank Michael Gastegger for discussion regarding the training process of ML models. B.L. is a recipient of a DOC Fellowships of the Austrian Academy of Sciences (ÖAW) at the Institute for Molecular Modeling and Simulation at the University of Natural Resources and Life Sciences, Vienna (grant no. 25743). The project was further supported by the Austrian Science Fund, FWF doctoral program BioToP (W1224) and the Erwin-Schrödinger project J 4522-N (J.W.). The computational results presented have been achieved in part using the Vienna Scientific Cluster (VSC).

Supplementary Information for:

BuRNN: Buffer Region Neural Network Approach for Polarizable-Embedding Neural Network / Molecular Mechanics Simulations

*Bettina Lier,[#a] Peter Poliak,[#a,b] Philipp Marquetand,[c] Julia Westermayr[*d] and Chris Oostenbrink[*a]*

a.  Institute for Molecular Modeling and Simulation, Department of Material Sciences and Process Engineering, University of Natural Resources and Life Sciences, Vienna. Muthgasse 18, 1190 Vienna Austria

b.  Department of Chemical Physics, Institute of Physical Chemistry and Chemical Physics, Faculty of Chemical and Food Technology, Slovak University of Technology in Bratislava, Radlinského 9, 812 37 Bratislava, Slovakia

c.  Institute of Theoretical Chemistry, University of Vienna, Währingerstraße 17, 1090 Vienna, Austria

d.  Department of Chemistry, University of Warwick, Gibbet Hill Road, CV4 7AL Coventry, UK.

*# Authors contributed equally*

*\* Corresponding authors: julia.westermayr@warwick.ac.uk and*

*chris.oostenbrink@boku.ac.at*

# Contents

# S1 Computational Details

## S1.1 Quantum Chemical Calculations

Total electronic energies and gradients of the training data points were all calculated in the Gaussian 16 program package[1] using spin-unrestricted density functional theory (DFT) in the gas phase. The iron-containing structures were calculated in the sextet state with a net charge of +3. Based on the benchmark calculations of iron complexes,[2] the OPTX exchange[3] with PBE correlation functional[4] (OPBE) were used in def2SVP basis set.[5] The self-consistent field (SCF) procedure was performed using the 'yqc' option. The energy minimisations were carried out by the default Berny optimisation algorithm. The spin contamination of the sextet species was below $10^{-5}$ a.u.. The use of the input orientation was enforced using the 'nosymm' keyword. Partial atomic charges were obtained from single-point calculations on the inner+buffer region by fitting to the electrostatic potential using the Merz-Singh-Kollman scheme.[1, 6-8] The partial atomic charges in the water droplet of 1.0 nm diameter calculated by this approach are very similar to the charges of the SPC water model, therefore further adjustment was not needed. Note that the fitting procedure bears the risk that the partial charges of the inner core are insensitive to the electrostatic potential further out. However, due to the large central charge on the iron ion, this problem does not occur. See Figure S1 for the distributions of the partial charges on Fe, O and H, in the training set, which are very stable. Average charges in the data set were 1.29, -0.73 and 0.41 for the iron, oxygen and hydrogen atom, respectively. Panel b shows the atomic charges as a function of the distance to $Fe^{3+}$. The outliers in the oxygen charges come from configurations in the training set in which the water molecules are oriented with their hydrogen atoms towards the $Fe^{3+}$. We note that use of gas-phase calculation for energies, gradients and partial charges is crucial as the buffer region in BuRNN interacts explicitly with the water in the outer region and the effect of, e.g., an implicit solvent would lead to double counting of solvation effects.

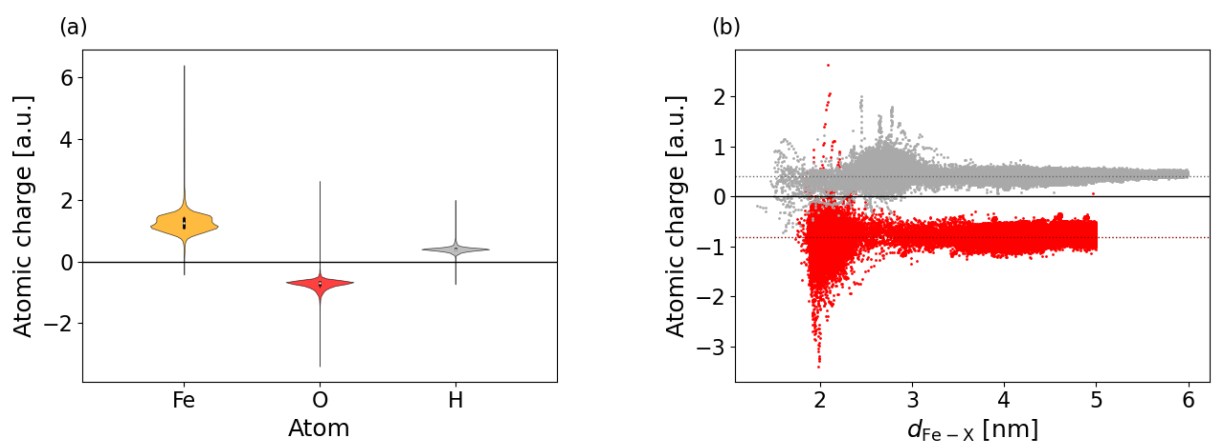

*Figure S 1: (a) Partial atomic charges of the data set and (b) their dependence on the distance to the Fe atom for oxygen (red) and hydrogen (grey). Dotted lines represent corresponding atomic charges in the SPC water model.*

## S1.2 Data set generation

The initial data set was generated from a molecular dynamics (MD) simulation (pure molecular mechanics description) set up as described in the Methods section S1.4. From an 8-ns MD simulation, snapshots were selected every 4 ps. Water molecules with the oxygen atom further than 0.5 nm from the iron atom were discarded, yielding Sa total of 2,000 configurations, for which single point QM calculations of the inner+buffer and for the buffer region alone were performed in gas-phase, as further described in section S1.1. After training on the initial training set, short MD simulations and optimizations with the BuRNN approach were performed on a system containing inner, buffer, and outer regions. These simulations did not lead to stable hexa-coordinated water configurations, but another 100 configurations were generated. To enhance accuracy close to the minimum, the 21 lowest energy configurations from the BuRNN simulation were extracted and minimized at the QM level. From these minimization trajectories, about 4,000 new configurations were extracted. The initial model was retrained on the updated data set, which led to much more stable BuRNN simulations, but we still observed irregular behaviour in the coordination sphere and the boundary region and large model deviations during simulations. Therefore, we postprocessed all configurations by a) applying SHAKE to ensure that the conformations of the water molecules adhered to the SPC configuration and b) removing any water molecules that in the QM optimization moved outside of the 0.5 nm cutoff. This added the latest 4,000 configurations a second time to the training set.

The NNs that were trained after this step led to stable simulations in the BuRNN approach, with differences between two independent NNs on the order of 10 kJ/mol. We next generated one additional set of conformations a) by picking the conformations for which the disagreement between the two independent NNs was largest during a BuRNN simulation (1,000 configurations), b) by running a BuRNN MD simulation with applying a biasing potential derived from the disagreement between the two independent NNs to sample the undertrained conformational space as

$$V_{\text{bias}} = \begin{cases} \frac{1}{4} K_{bias} \left[ \left( V_{\mathbb{I}+\Delta\mathbb{B}}^{NN(A)} - V_{\mathbb{I}+\Delta\mathbb{B}}^{NN(B)} \right)^2 - V_{val,thresh}^2 \right]^2, & V_{\mathbb{I}+\Delta\mathbb{B}}^{NN(A)} - V_{\mathbb{I}+\Delta\mathbb{B}}^{NN(B)} > V_{thresh} \\ 0, & V_{\mathbb{I}+\Delta\mathbb{B}}^{NN(A)} - V_{\mathbb{I}+\Delta\mathbb{B}}^{NN(B)} \leq V_{thresh} \end{cases} \quad \text{S1}$$

where $K_{bias}$ was set to –0.01 kJ$^{-3}$ mol$^3$, $V_{thresh}$ to 1.0 kJ/mol$^{-1}$ and $V_{\mathbb{I}+\Delta\mathbb{B}}^{NN(B)}$ is a predicted energy from the second NN model (500 configurations), c) selecting conformations from a simulation in which one water molecule was artificially pulled away from the $Fe^{3+}$ (300 configurations). This was done to ensure that the NNs learned energies and partial charges with a water molecule between the first and second solvation shell as well. Configurations were compared by energies, forces and spin contamination and obvious outliers were discarded leading to an overall training set of about 11,000 configurations, of which 9,500 were used for training and the rest for testing.

We emphasize that the procedure outlined above was the result of a continuous process to obtain a sufficiently large training set, such that we could test the BuRNN approach and not the training. For any further applications we are confident that generation of an initial training set that is SHAKEN and filtered up to the buffer region, followed by an adaptive sampling scheme to generate additional conformations as needed will be sufficient.

### S1.3 Neural Networks

### S1.3.1. Neural network training

Deep NNs used in this work are based on the continuous-filter convolutional NN SchNet,[9, 10] which is a message-passing NN that learns the molecular descriptor in addition to its relation to target properties. In this work, we use SchNet to train on the interaction energies, $V_{\mathbb{I}+\Delta\mathbb{B}}^{NN}$ (see equations 7 and 8 in the main text) and the corresponding interaction forces, $\boldsymbol{F}_{\mathbb{I}+\Delta\mathbb{B}}^{NN}$, calculated as derivatives of the NN potentials:

$$\boldsymbol{F}_{\mathbb{I}+\Delta\mathbb{B}}^{NN} = -\frac{\partial V_{\mathbb{I}+\Delta\mathbb{B}}^{NN}}{\partial \boldsymbol{R}}. \quad \text{S2}$$

$\boldsymbol{R}$ denotes the atomic positions of all atoms within the QM region (inner and buffer region) and we used a shorthand notation to indicate the derivatives with respect to each of the elements of the vector.

The loss function used for training contains both terms, i.e., interaction energies and interaction forces (see equations S2-S4). The training set contained outliers, i.e., unfavourable structures, e.g., 1-4 coordinated aqua-Fe-complexes, with large energies and forces (but not necessarily large interaction energies and interaction forces). To allow a training of a diverse data set including such "outliers", we used a variant of the smooth $L_1$ loss function, which switches from $L_2$ to $L_1$ whenever a data point is deemed to be an outlier. The loss functions used for training energies and forces are given below in equations S3 and S4.

$$L_2 = t \left\| V_{\mathbb{I}+\Delta\mathbb{B}}^{NN} - \left(V_{\mathbb{I}+\mathbb{B}}^{QM} - V_{\mathbb{B}}^{QM}\right) \right\|^2 + (1-t) \left\| -\frac{\partial V_{\mathbb{I}+\Delta\mathbb{B}}^{NN}}{\partial \boldsymbol{R}} + \left(\frac{\partial V_{\mathbb{I}+\mathbb{B}}^{QM}}{\partial \boldsymbol{R}} - \frac{\partial V_{\mathbb{B}}^{QM}}{\partial \boldsymbol{R}}\right) \right\|^2 \quad \text{S3}$$

$$L_1 = t \left| V_{\mathbb{I}+\Delta\mathbb{B}}^{NN} - \left(V_{\mathbb{I}+\mathbb{B}}^{QM} - V_{\mathbb{B}}^{QM}\right) \right| + (1-t) \left| -\frac{\partial V_{\mathbb{I}+\Delta\mathbb{B}}^{NN}}{\partial \boldsymbol{R}} + \left(\frac{\partial V_{\mathbb{I}+\mathbb{B}}^{QM}}{\partial \boldsymbol{R}} - \frac{\partial V_{\mathbb{B}}^{QM}}{\partial \boldsymbol{R}}\right) \right| \quad \text{S4}$$

The tradeoff, *t*, is used to weigh forces and energies during training and was sampled between 0.001 and 100. The optimal value was found to be 0.05. Note that we train forces and not gradients, hence the negative sign.

*Atomic partial charges*

To describe atomic partial charges, another SchNet model was trained on the full QM region (inner + buffer region). This calculation could be performed on the already converged orbital coefficients of the inner+buffer region, making the additional computational costs negligible and keeping the QM calculations at two.

To allow training of atomic partial charges, the output modules of SchNet had to be adapted such that all atomic partial charges could be learned by one NN. Each atom described in its chemical and structural environment by the NN gave rise to a corresponding atomic partial charge. Therefore, the last pooling layer, which usually sums or averages over all atomic contributions for a target property, was removed. During the simulations, the predicted charges were adjusted to sum up to +3 exactly by homogeneously distributing any charge deficit or surplus (typically 5%) over all particles in the inner and buffer regions.[11]

In contrast to interaction energies and interaction forces, the atomic partial charges, $q_a$, with *a* indicating an atom in the whole system containing $N_a$ atoms, were modelled using an $L_2$ loss:

$$L_2^{charges} = \sum_a^{N_a} \left\| q_{a,\mathbb{I}+\mathbb{B}}^{NN} - q_{a,\mathbb{I}+\mathbb{B}}^{QM} \right\|^2. \quad \text{S5}$$

Using a smooth $L_1$ function instead of the $L_2$ loss did not improve training. As can be seen, multiple values are treated in one NN.

S1.3.2 Hyperparameter optimization

Hyperparameters were assessed for both, interaction energy and interaction force models and atomic partial charge models separately. A random grid search was applied to obtain optimal model parameters. Besides default parameters, we use a cutoff of 1.0 nm and a batch size of 8. The network parameters were validated on a random grid using a training:validation:test split of approximately 8:1:1 and an intermediate training set of 7,900 data points. As SchNet is a message passing neural network (NN), which automatically generates a tailored representation for a given system, it can be seen as a connection of two NNs. The size of the network that models the descriptor based on the structural inputs and elemental charges is defined by 6 interaction layers, 256 features to represent the atoms and 50 Gaussian functions. The number of Gaussian functions was set to 100 for modelling atomic partial charges. The interaction layers were sampled from 4-8, the features from 128-512, and the number of Gaussian functions placed on each atom from 25-200. The learning rate was varied between 0.001 and 0.00001, whereas the default of 0.0001 was most appropriate for training the final data set using 9,500 data points for training. A larger learning rate up to 0.001 was used when training smaller training set sizes. The hidden layers to map the descriptor to the output energies was kept at 3 with more layers not improving training.

S1.3.2. Neural network accuracy

Training directly on interaction energies and interaction forces has several advantages compared to training on the separate terms arising from the inner+buffer region and the buffer region alone. On one hand, our setup allows for higher accuracy and consequently, better data efficiency, as the interaction energies span a smaller energy window than the total energies of the inner+buffer region and the buffer region. While the mean absolute error (MAE) on total energies and forces for models trained on 9,500 data points is in the range of 500 kJ/mol and 5,000 kJ/mol/nm, respectively (resulting in an error of around 700 kJ/mol and 7,000 kJ/mol/nm for interaction energies and interaction forces, respectively), training directly on interaction energies and interaction forces leads to errors about 100-1000 times smaller, i.e., $1.7 \pm 0.3$ kJ/mol and $8.4 \pm 0.4$ kJ/mol/nm, respectively. The MAEs and root mean squared errors (RMSE) on a holdout test set for models trained on interaction energies and interaction forces are shown in Table S1. MAEs and RMSEs of 5 independently trained models for partial charges on 9,500 data points are shown in Table S2.

*Table S1: Mean absolute error (MAE) and root mean squared error (RMSE) for energies and forces of 5 models trained on 9,500 data points corresponding to Fig. S1 (a-d).*

| Neural Network Model | Energy MAE (RMSE) [kJ/mol] | Forces MAE (RMSE) [kJ/mol/nm] |
|---|---|---|
| **Model 1** | 1.73 (8.32) | 8.35 (21.29) |
| **Model 2** | 2.17 (7.94) | 9.05 (26.16) |
| **Model 3** | 1.51 (8.68) | 7.94 (23.02) |
| **Model 4** | 1.60 (5.55) | 8.68 (24.17) |
| **Model 5** | 1.46 (4.63) | 8.08 (19.65) |

*Table S2: Mean absolute error (MAE) and root mean squared error (RMSE) for charges of 5 models trained on 9,500 data points corresponding to Fig. S1(e-f).*

| Neural Network Model | Charges MAE (RMSE) [a.u.] |
|---|---|
| **Model 1** | 0.026 (0.059) |
| **Model 2** | 0.027 (0.060) |
| **Model 3** | 0.026 (0.057) |
| **Model 4** | 0.027 (0.057) |
| **Model 5** | 0.027 (0.059) |

To assess whether models can learn interaction energies, interaction forces, and atomic partial charges properly, we computed learning curves, which show the mean absolute error of several individually trained models for a given training set size in logarithmic scale. For proper learning, a linear decline is expected for the loss function on a log-log scale. The number of models that were trained for a given training set size was chosen such that the standard deviation of the MAE was within 1 kJ/mol (less than 10 meV).

The learning curves, in which each point shows the mean and the standard deviation of the mean absolute error (MAE) for 4-20 models trained on the interaction energy, interaction forces, and atomic partial charges, are shown in Figure S2, panels a, c, and e, respectively. The corresponding scatter plots of 5 models trained on interaction energies and interaction forces using the largest training set size are illustrated in panels b and d, respectively. For charges, only one model is used for dynamics, hence only one model is shown in panel f. The scatter plots are shown for the test set. In each plot, the systems for which the QM calculations are performed are indicated at the bottom. As can be seen from the scatter plots, the model learns interaction energies and forces accurately. Except for a few data points with larger errors (that scatter stronger), model predictions almost perfectly match the reference values. Data points with larger errors are related to energetically unfavourable structures that were visited mainly during the last adaptive sampling run, mostly during dynamics with BuRNN on long time scales. As can be seen from the learning curves, all models learn properly, as a linear relation is observed when plotting the data on a log-log scale for the number of data points used for training and the MAEs of models. As is visible in panel f, the charge models scatter more strongly than the models for interaction energies and interaction forces. The learning curve shows that more data points could still improve the accuracy if needed. BuRNN simulations on long time scales were robust and led to smooth transitions in the forces between the buffer and outer regions (Figure S2), indicating that the accuracy of the charge predictions is sufficient for production runs.

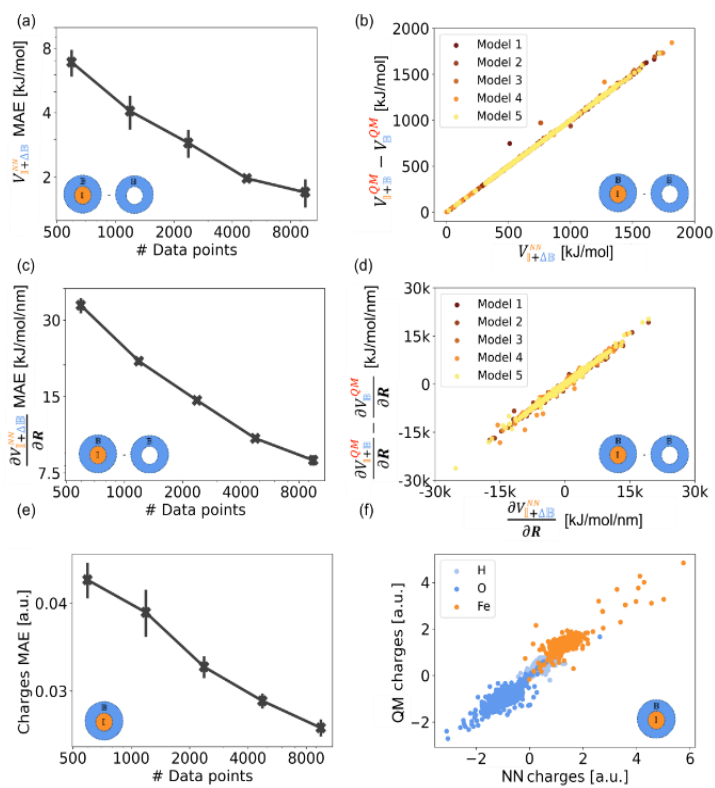

*Figure S2: Learning curves for (a) interaction energies, (c) interaction forces, and (e) atomic partial charges that show the mean of 4-20 models trained on a given training set size including their standard deviation. Learning curves are shown in a logarithmic scale. Scatter plots for (b) interaction energies, (d) interaction forces, and (f) atomic partial charges of models trained on the largest training set size are shown. For adaptive sampling, at least two NN models are used for energies and forces, while only one model is used for atomic partial charges, hence only one model is shown. The interaction or region that is modelled is indicated at the bottom of each panel.*

To investigate the performance of the model on larger inner regions, we additionally created a data set using a larger inner region with a cutoff of 2.5 nm from the central $Fe^{3+}$ ion. We picked data points that were from SHAKEN simulations out of the original data set and 6211 calculations converged. We then trained neural networks on 5300 data points and used 300 and the rest for validation and testing, respectively. The errors we get when comparing two independently trained models using the smooth L1 loss function (section S1.3.1) are comparable to the models trained on similar amounts of data using an inner region that comprises the $Fe^{3+}$ ion only.

The MAE (RMSE) for the models trained on the smaller inner region are 1.97 kJ/mol (4.73 kJ/mol) for energies and 12.1 kJ/mol/nm (42.3 kJ/mol/nm) MAE (RMSE) for forces. The models trained on the larger inner region have an MAE (RMSE) of 1.97 kJ/mol (7.9 kJ/mol) for energies and of 10.3 kJ/mol/nm (22.9 kJ/mol/nm) for forces, which is in both cases remarkably small.

These models were further used for BuRNN simulations. The radial distribution function is plotted for comparison in Figure S4 below. Results are comparable to those obtained from BuRNN with a smaller inner region.

## S1.4 Molecular Dynamics Simulations

All molecular dynamics simulations were performed with a modified version of the GROMOS simulation package,[12, 13] with a direct interface to SchNetPack[10] modules using the pybind11 library.[14] As the code is deeply integrated into GROMOS, it will be part of the next release and will become freely available at www.gromos.net. Until then, the code is available upon request. Example input files to run a BuRNN simulation are provided.[15]

Simulations were performed in a periodic cubic box with box-edge lengths of 2.91 nm, containing one $Fe^{3+}$ ion, and 786 SPC water molecules.[16] Temperature was maintained at 298 K using the Nosé Hoover chains scheme[17] by coupling the centre-of-mass motion and internal/rotational degrees of freedom to two separate temperature baths with coupling time of 0.1 ps and four chains. Bond lengths and angles in SPC water were constrained using the SHAKE algorithm.[18] Unless stated otherwise, a timestep of 2 fs was used. Lennard-Jones parameters of the $Fe^{3+}$ atom were taken from the IOD parameter set of Li et al.[19] and the effective charge in classical simulations was set to +3.0. An additional set of simulations was performed with the 12-6-4 Lennard-Jones potential, in conjunction with the SPC/E water model, to account for the ion-induced dipole moment of the solvent, as derived by Li and Merz.[19] For the nonbonded MM interactions we applied a cutoff of 1.4 nm based on the distances between the heavy atoms (charge-group based cutoff) with a reaction-field of relative permittivity of 61 to account for a homogeneous medium outside the cutoff.[20] The pairlist was updated every step.

The BuRNN buffer region was created by applying a charge-group based cut-off of 0.5 nm from the central ion. Energies and forces were obtained from the previously trained Schnet models. We simultaneously used two models trained on the same data sets with different training and validation splits. The second model was used to validate the first model by comparing the energy predictions on-the-fly. Their mutual disagreement was used to pick snapshots for the next round of adaptive learning. Our implementation allows to monitor this quantity throughout the simulation. Partial atomic charges were obtained from another, adapted atomistic Schnet model[21] and applied to the $Fe^{3+}$ ion and the water molecules inside the buffer region. To avoid double-counting of the water-water interactions within the buffer, their standard MM parameters and partial charges were used, as the polarization effects are added through the NN. The NN attributed charges were used only for the Coulombic interactions of the inner and buffer regions with the outer region. MM and BuRNN simulations were performed for 10 ns.

Simulations with a larger inner region, including the six coordinating waters were performed for 1 ns. As the current code does not allow for particles to change from the inner to the buffer region, additional half-harmonic attractive restraints were added to these waters. This ensures that the inner water molecules remain in the inner region and do not diffuse into the buffer or outer regions.

QM/MM simulations were performed using the central $Fe(H_2O)_6^{3+}$ complex as the QM region and the remaining water molecules as classical SPC water. QM calculations were performed by Gaussian 16, at the same functional and basis set as described above. Simulations were performed using both mechanical and electrostatic embedding schemes, with nonpolar nonbonded interactions between the QM and MM regions described using the same force field parameters as described above. All water bonds and angles remained constrained using the SHAKE algorithm, allowing for a timestep of 2 fs and a total simulation time of 20 ps. In the mechanical embedding scheme, the coulombic interactions between the QM region and MM region were calculated using charges from the QM calculation every step. In the electrostatic embedding scheme, the MM charges up to 0.8 nm from the inner region were included in the SCF calculation.

The coordination number of the iron ion was determined using a 0.25 nm cutoff, based on an analysis of initial radial distribution functions. Hydrogen bonds were determined using a geometric criterion and were considered present if any hydrogen atom was within 0.25 nm of the oxygen atom of a neighbouring water molecule, while the O-H··O angle was at least 135°. Diffusion of the iron ion was determined using the Einstein equation by applying a linear fit to the mean square displacement over timescales of 0 – 0.5 ns.

## S2 BuRNN simulations

### S2.1 BuRNN forces

Figure S3 shows the profiles of the net forces on water molecules in panel a and their radial components in panel d. The net forces are smooth at the boundary (0.5 nm), which can be attributed to the fact that at the boundary the interaction with the inner region is small in comparison to the rest of the system. Contrary, the atom-wise contributions (panels c and f) exhibit jumps at the boundary, reflecting the transition from an electrostatic interaction based on localized point charges to a quantum mechanical interaction, mediated by delocalized electron densities. For complete molecules, these jumps disappear. Panels b and e show the contribution of the NN to the forces (either directly in $\boldsymbol{F}_{\mathbb{I}+\Delta\mathbb{B}}^{NN}$ or via NN-prediced partial charges in $\boldsymbol{F}_{\mathbb{B},\mathbb{O}}^{MM}$ and $\boldsymbol{F}_{\mathbb{I},\mathbb{O}}^{MM}$) is displayed. Within the buffer region, the forces reflect the observations in the radial distribution functions.

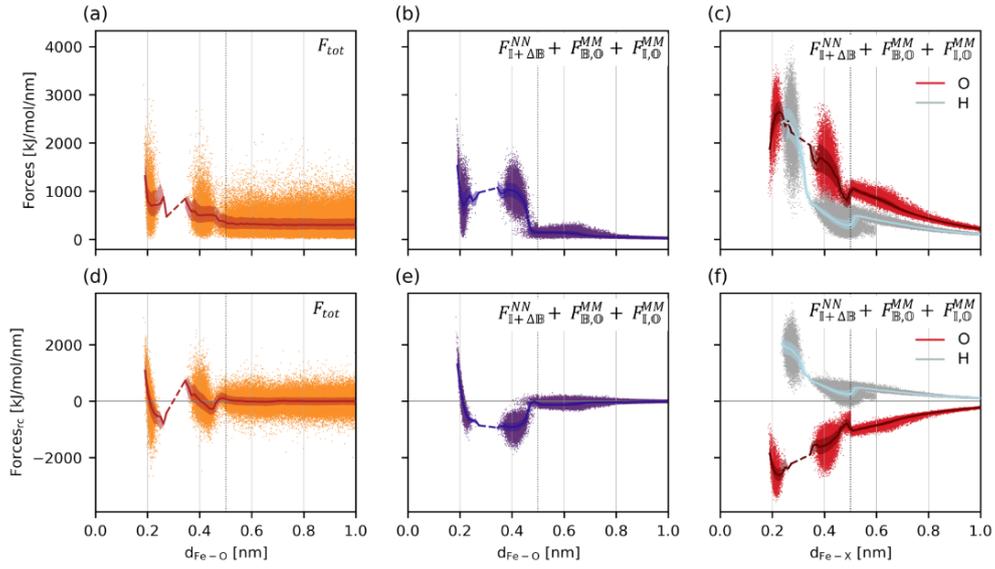

*Figure S3: Forces as a function of the distance to the iron atom (dots). The line represents the median and the band corresponds to the 2<sup>nd</sup> and 3<sup>rd</sup> quartile. Net forces on water molecules (**a**), contribution of the NN treatment of the inner and buffer region ($F^{NN}_{\mathbb{I}+\Delta\mathbb{B}} + F^{MM}_{\mathbb{B},\mathbb{O}} + F^{MM}_{\mathbb{I},\mathbb{O}}$) to the net forces on the water molecules (**b**) and on the oxygen (red) and hydrogen (green) atoms (**c**). Panels **d**, **e** and **f** show radial components of the respective forces. Negative value means attraction.*

## S2.2 Observed structure

Figure S4 shows the radial distribution functions for all simulations described in this work. As can be seen from the red and orange lines, BuRNN simulations with different inner regions are comparable. The orange curve is obtained from an inner region comprising the $Fe^{3+}$ ion and the red curve is obtained from an inner region comprising the $Fe^{3+}$ ion and the first solvation shell (a cutoff of 2.5 nm was chosen). Note that the peak corresponding to the second solvation shell resembles the one observed with QM/MM using mechanical embedding (abbreviated as ME in brackets), while for QM/MM with electrostatic embedding (abbreviated as EE in brackets) this peak is more similar to the MM only simulations. When using the 12-6-4 Lennard-Jones potential, an additional small peak between the first and second solvation shell is observed, corresponding to a water molecule that is in rapid exchange with the first shell. We have also performed simulations with different charge distributions on the inner and buffer regions in an overlay. Using partial charges from the force field for the inner and buffer regions leads to an artefact at the buffer region cutoff at 0.5 nm. Using (NN-predicted) charges from an additional implicit solvent QM calculation reduces, but does not remove this artefact. In these QM calculations we used the SMD variant of the integral-equation-formalism polarizable continuum model.[8] With the charges that are derived from gas-phase QM calculations, we obtain a smooth transition at 0.5 nm.

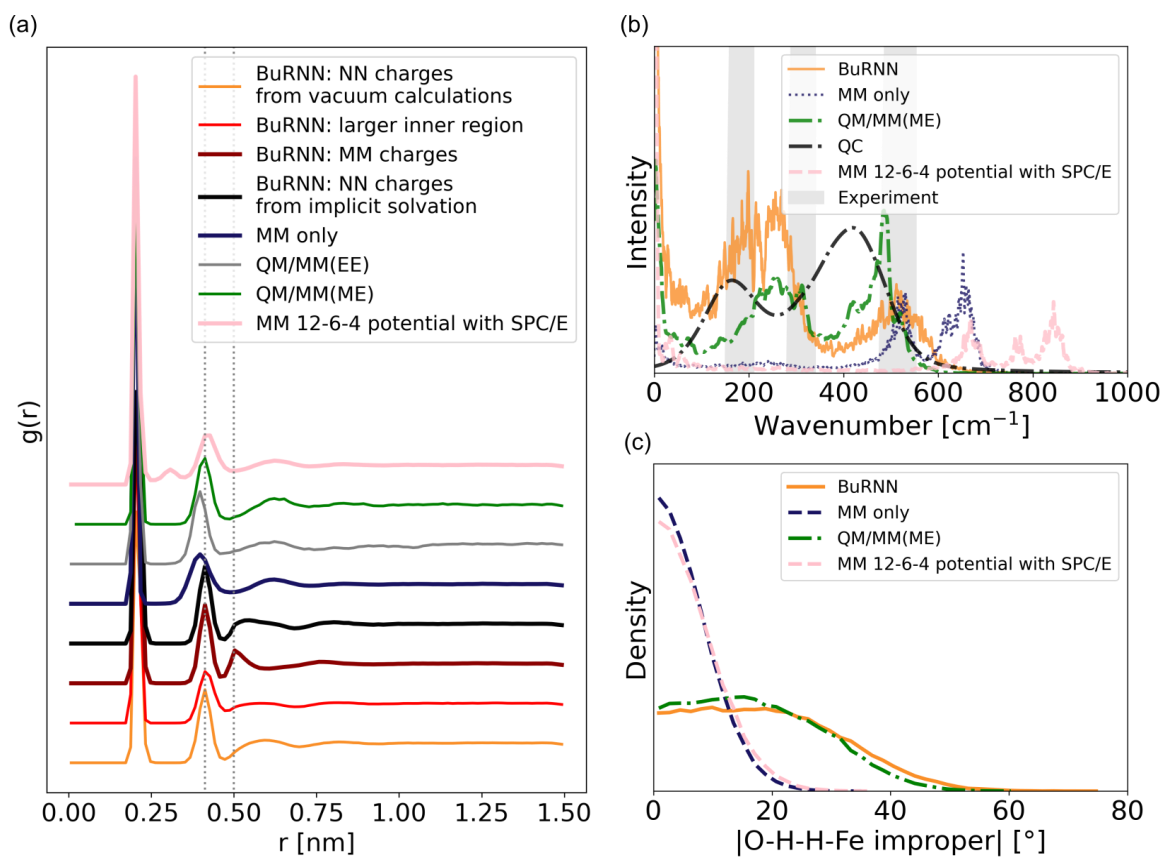

*Figure S4: a) Iron-oxygen radial distribution function for different simulations settings. The blue curve shows the radial distribution function for a purely classical simulation (MM only), the orange (red) one for a regular BuRNN simulation (with a larger inner region), in which the partial charges for the inner and buffer region were determined from the QM calculations in vacuum. The black and dark red curves show the radial distribution function for a simulation in which partial charges for the inner and buffer region were estimated from QM calculations in an implicit solvent and assigned according to the classical force field, respectively. Grey and green lines show QM/MM simulations with density functional theory, once using an electrostatic embedding (EE) and once an mechanical embedding (EE), respectively. The pink line shows the radial distribution function obtained with MM 12-6-4 potential with SPC/E. Dotted lines indicated the second BuRNN peak and the cutoff used for the buffer region. Dashed lines indicate the second BuRNN peak and the cutoff used to define the buffer region b) Comparison of power spectra. QC refers to a quantum chemistry calculation of $Fe(H_2O)_{21}$ with the reference method used to generate the training set. Only modes related mostly to the Fe-O bond lengths are shown. The bands were created with a half-peak width of 100 $cm^{-1}$ and for broadening a 1:1 mixture of Gaussian and Lorentzian functions were used. c) Probability distribution of the O-H-H-Fe improper dihedral angle obtained from BuRNN, MM only, QM/MM with EE and the MM 12-6-4 potential with SPC/E simulations.*

During the 10-ns BuRNN simulation the six coordinating water molecules did not exchange spontaneously. In the MM only description, a seventh water molecule very rarely moves within 0.25 nm of the central iron. In BuRNN, few individual water molecules occasionally vibrate beyond the 0.25 nm cutoff, leading to coordination numbers smaller than 6. No complete exchanges of the coordinating water molecules were observed in these simulations. In contrast, out of the total number of 786 water molecules, all visit the buffer region at least once during the 10-ns simulation. In total, 8,847 distinct visits to the buffer region were monitored (not counting the six coordinating water molecules), with an average lifetime of 14.4 ps. 8,579 visits had a lifetime of 5 ps or less; 268 visits to the buffer region had lifetimes longer than 5 ps, up to a maximum of 2,240 ps.

To assess differences in pure MM and BuRNN simulations, Figure 3 in the main text summarizes the geometries that are observed of the hexa-aqua iron complex. Figure S4 shows additional data for the simulations using QM/MM with mechanical embedding and for the classical simulations with the 12-6-4 Lennard-Jones potential. Note that consistently, the classical descriptions show an improper dihedral distribution that is centered around 0°, while the QM/MM and BuRNN simulations show wider distributions with an average value of 19 to 20°. While the distributions of distances are quite comparable in all simulations, the dynamics of the Fe-O bond is rather distinct. The power spectrum in Figure 3d and Figure S4c was calculated as the Fourier transform of the autocorrelation function of the Fe-O bond lengths, as obtained in simulations in which these distances were recorded at every timestep for 100 ps (20 ps for the QM/MM simulations). Simulations using timesteps of 0.5 fs and of 2 fs are indistinguishable, with the period of the fastest vibrational motion around 50 to 60 fs, confirming that a timestep of 2 fs is suitable. The spectra obtained with BuRNN are most similar to those obtained from QM/MM calculations and are very different from both classical descriptions. In addition, we include the vibrational modes obtained from a quantum chemical calculation with the reference method using the optimized geometry and rigid-rotor/harmonic-oscillator approximation for $Fe(H_2O)_{21}$ in Figure S4c. Only modes related mostly to the Fe-O bond lengths are shown. The band is plotted as a superposition of broadened spectral lines of these modes using mixed Gaussian/Lorentzian functions (50:50) with a half-peak width of 100 $cm^{-1}$ to resemble the shape of experimental spectra. These peaks are better aligned with BuRNN than with a pure MM simulation. The remaining differences between the QC spectrum and the observed frequencies in the power spectra can be explained by anharmonicities arising from the larger environment in the simulation. Moreover, the BuRNN spectra are in closer agreement with experimentally observed bands of 180 $cm^{-1}$, 310 $cm^{-1}$ and 500 $cm^{-1}$.[22-24]

S2.3 BuRNN is robust with respect to higher temperatures

During BuRNN simulations at elevated temperatures, the disagreement between two NNs was monitored. The difference between the two NNs remained at an average value of $-0.39 \pm 0.02$ kJ/mol with a standard deviation that increased from 1.17 kJ/mol at 300 K to 1.45 kJ/mol at 400 K, which is still very small and stays below chemical accuracy of 1 kcal/mol. Figure S5 shows the radial distribution functions and coordination numbers of 1-ns simulations at different temperatures. While the number of configurations in which a water molecule moves outside of the 0.25 nm cutoff increases, no complete exchanges of water molecules are observed at any of these temperatures.

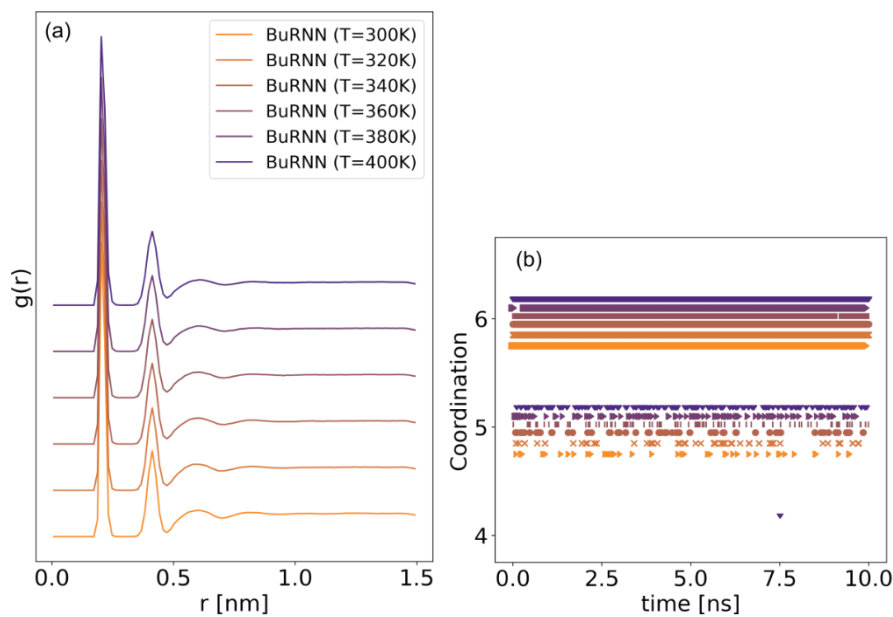

*Figure S5: Iron-oxygen radial distribution functions for BuRNN simulations at different temperatures (left) and time series of the number of water molecules within 0.25 nm of the Fe3+ ion (with a small offset to visualize multiple simulations (right). Due to the size of the marker, the bars at a coordination of 6 seem continuous, but a small number of configurations is observed for which only 5 water molecules are within 0.25 nm.*

Regarding simulations at which we pulled a water molecule away from Fe, we evaluated NN differences every 0.05 ps. The disagreement was found to be in the range of –0.68 ± 0.10 kJ/mol with a standard deviation of 1.47 kJ/mol, suggesting that the NNs are quite robustly sampling penta-coordinated iron complexes as well.

*Movie S1: The file movie_S1.mp4 shows the spontaneous binding of a water molecule (cyan) to a pentacoordinated $Fe^{3+}$-water complex. The configuration switches from a trigonal bipyramidal arrangement to an octahedral complex in the process. $Fe^{3+}$ as orange sphere, coordinating waters in sticks, buffer waters in blue, water in the outer region in grey.*